\documentclass{iopart}
\usepackage{setstack}
\usepackage{graphicx}

\newcommand{\dyad}[1]{\bi{#1}}

\begin{document}
\title{Electromagnetic multipole theory for optical nanomaterials}

\author{P. Grahn, A. Shevchenko, and M. Kaivola}
\address{Department of Applied Physics, Aalto University, P.O. Box 13500, FI-00076 Aalto, Finland}

\begin{abstract}
Optical properties of natural or designed materials are determined by the electromagnetic multipole moments that light can excite in the constituent particles. In this work we present an approach to calculate the multipole excitations in arbitrary arrays of nanoscatterers in a dielectric host medium. We introduce a simple and illustrative multipole decomposition of the electric currents excited in the scatterers and link this decomposition to the classical multipole expansion of the scattered field. In particular, we find that completely different multipoles can produce identical scattered fields. The presented multipole theory can be used as a basis for the design and characterization of optical nanomaterials.
\end{abstract}

\section{Introduction}
The classical electromagnetic multipole expansion \cite{JacksonBook} is a powerful tool for analyzing the electric and magnetic fields created by spatially localized electric charges and currents. Irrespective of the complexity of the charge and current distributions, the fields produced by them can be represented as a superposition of the fields created by a corresponding set of point multipoles. This correspondence provides a common basis for characterizing the fields radiated by localized charge and current excitations in arbitrary configurations.

In optics, the multipole expansion is well suited to describe the scattering of optical fields by small objects. Usually, if the wavelength of the field is large compared to the size of the object, the scattering is described mainly by the lowest-order multipole, the electric dipole, while the contributions from all higher-order multipoles are considered as mere perturbations. Recently, it has been shown that in specifically designed optical nanomaterials \cite{CaiBook}, such as metamaterials, the contribution of the magnetic dipole \cite{Yen05032004} and the electric quadrupole \cite{Cho2008} excitation to the scattering by the material's constituents can be made significant, which substantially affects the optical properties of the material and lead to extraordinary phenomena, such as negative refraction \cite{Smith1999}. In certain materials, higher-order multipoles can even completely overshadow the electric dipole contribution \cite{Metadimer}. Thus, it is clear that higher-order multipoles have to be taken into account when evaluating the macroscopic electromagnetic characteristics of such materials \cite{MultiMetaColl}.

In order to create a material with prescribed optical properties, one should select the elementary unit of the material (often called the meta-atom) and optimize its scattering characteristics through adjusting the design. For an individual particle this can be done by numerically solving the Maxwell equations for the scattered field and applying the multipole expansion to determine the multipoles contributing to the scattering \cite{Muhlig201164}. However, in a material composed of a large number of such elementary units, each particle interacts with the fields scattered by the other particles, which can significantly modify the amplitudes and phases of the excited multipoles. Since at each point in the material the field is a superposition that contains the fields scattered by all the particles, the approach developed for individual particles \cite{Muhlig201164} can no longer be used, and one should perform the multipole decomposition on the excited electric currents in each of the particles individually. Previously, this decomposition has been introduced only for a single localized electric current distribution in vacuum \cite{JacksonBook} and, therefore, a multipole theory suitable for analysis of nanoscatterers in an array and in an arbitrary homogeneous dielectric host medium was missing. In this work, we introduce such a theory. Our multipole expansion approach is particularly suitable for direct numerical implementation.

The geometry of the scatterer determines the types of electric current modes that can be excited in it by light. For the description of these modes, we propose a set of orthogonal electric current multipoles, which are composed of elementary point currents in simple configurations. Each element of the resulting current multipole tensor reflects the strength of one of these configurations, which enables one to visualize the real electric current modes that will be excited in a scatterer.
We complete our theory by deriving expressions that relate the elements of the proposed electric current multipole tensor to the classical multipole expansion coefficients. In particular, these expressions reveal i) perfectly dark multipole modes that do not create any electromagnetic field, and ii) electric dipole radiation produced by electric currents with zero net electric dipole moment. These findings provide us with additional freedom in the choice of the particle geometry.

In section~\ref{Sec2}, the classical multipole expansion is adjusted to describe the electromagnetic field scattered by individual nanoparticles and nanoparticle arrays embedded in a dielectric host medium. In section~\ref{Sec3}, we map the coefficients in the multipole expansion to the electromagnetic fields created by sets of sub-wavelength current elements. Explicit mapping relations are derived up to the orders of electric octupole and magnetic quadrupole, which both describe third-order excitations in the multipole hierarchy \cite{RaabBook}. In section~\ref{Sec4}, we summarize our results.

\section{The multipole expansion of the scattered field\label{Sec2}}

We consider a monochromatic plane wave, with the electric field amplitude $E_0$, angular frequency $\omega$ and wave vector $\bi{k}$, incident on a particle in an otherwise homogeneous lossless dielectric medium. In general, the scattered electromagnetic field can be written in spherical coordinates in the form of the following multipole expansion \cite{JacksonBook}
\begin{eqnarray}\label{multiE}\nonumber\fl
\bi{E}_{\rm{s}}(r,\theta,\phi) = E_0\sum_{l=1}^{\infty}\sum_{m=-l}^l \rmi^l[\pi (2l+1)]^{1/2}
\Big\{\frac{1}{k}a_{\rm{E}}(l,m)\nabla\times \big[h_l^{(1)}(kr)\bi{X}_{lm}(\theta,\phi)\big]\\
+ a_{\rm{M}}(l,m)h_l^{(1)}(kr)\bi{X}_{lm}(\theta,\phi)\Big\},\\\label{multiH}\nonumber\fl
\bi{H}_{\rm{s}}(r,\theta,\phi) = \frac{E_0}{\eta}\sum_{l=1}^{\infty}\sum_{m=-l}^l \rmi^{l-1}[\pi (2l+1)]^{1/2}
\Big\{\frac{1}{k}a_{\rm{M}}(l,m)\nabla\times\big[h_l^{(1)}(kr)\bi{X}_{lm}(\theta,\phi)\big]\\
+a_{\rm{E}}(l,m) h_l^{(1)}(kr)\bi{X}_{lm}(\theta,\phi)\Big\},
\end{eqnarray}
where $\bi{X}_{lm}$ and $h_l^{(1)}$ are the normalized vector spherical harmonics and the spherical Hankel functions of the first kind, respectively. The wave number $k$ is taken to be that in the surrounding dielectric which has an impedance $\eta$. We have chosen the normalization of the multipole expansion such that the expressions for the scattering and extinction cross sections are compact. The vector functions in the multipole expansion form a complete basis for representing the electromagnetic field outside an arbitrary localized source \cite{lambert:849}.

Let us now assume that the scattered electric field $\bi{E}_{\rm{s}}$ is known in the region surrounding the scatterer (e.g., calculated numerically). Using the orthogonality properties of the vector spherical harmonics $\bi{X}_{lm}$ and the scalar spherical harmonics $Y_{lm}$, we can calculate the multipole coefficients from the distribution of the scattered electric field on any spherical surface enclosing the scatterer as
\begin{eqnarray}\label{aeSphere}\fl
a_{\rm{E}}(l,m) &= \frac{(-\rmi)^{l+1}kr}{h_l^{(1)}(kr)E_0[\pi(2l+1)l(l+1)]^{1/2}}\int_0^{2\pi}\int_0^\pi Y^*_{lm}(\theta,\phi) \hat{\bi{r}}\cdot \bi{E}_{\rm{s}}(\bi{r})\sin\theta d\theta d\phi,\\\label{amSphere}\fl
a_{\rm{M}}(l,m) &= \frac{(-\rmi)^l}{h_l^{(1)}(kr)E_0[\pi(2l+1)]^{1/2}}\int_0^{2\pi}\int_0^\pi \bi{X}^*_{lm}(\theta,\phi)\cdot\bi{E}_{\rm{s}}(\bi{r})\sin\theta d\theta d\phi.
\end{eqnarray}
The above expressions for these coefficients can also be written in terms of the scattered magnetic field $\bi{H}_{\rm{s}}$. In this work, we use the magnetic coefficients expressed as
\begin{equation}\label{amSphereAlt}\fl
a_{\rm{M}}(l,m) = \frac{(-\rmi)^l\eta kr}{h_l^{(1)}(kr)E_0[\pi(2l+1)l(l+1)]^{1/2}}\int_0^{2\pi}\int_0^{\pi} Y^*_{lm}(\theta,\phi) \hat{\bi{r}}\cdot\bi{H}_{\rm{s}}(\bi{r})\sin\theta d\theta d\phi.
\end{equation}
For scattering by a symmetric particle, all coefficients of higher order in $l$ than some $l_{\rm{max}}$ can be made equal to zero by properly selecting the origin of the coordinate system. The value of $l_{\rm{max}}$ depends on both the size and the geometrical complexity of the particle.

Next, we assume that the particle belongs to a large array of similar particles. The multipole coefficients must then be deduced from the distributions of the electric current density in the particles. For a single particle, such a derivation is presented in \cite{JacksonBook} under the assumption that the surrounding medium is vacuum. In order to allow for a dielectric surrounding, we define a quantity
\begin{equation}\label{JS}
\bi{J}_{\rm{S}}(\bi{r}) = -\rmi\omega\epsilon_0\big[\epsilon_{\rm{r}}(\bi{r}) - \epsilon_{\rm{r,d}}\big]\bi{E}(\bi{r}),
\end{equation}
which we call the scattering current density. The electric field $\bi{E}=\bi{E}_{\rm{i}} + \sum_j\bi{E}_{\rm{s},j}$ contains both the incident field $\bi{E_{\rm{i}}}$ (the field in the dielectric in the absence of any scatterers) and the field $\bi{E}_{\rm{s},j}$ scattered by each particle $j$. In (\ref{JS}), $\epsilon_{\rm{r,d}}$ is the real-valued relative electric permittivity of the dielectric and $\epsilon_{\rm{r}}(\bi{r})$ the complex-valued relative electric permittivity at any coordinate $\bi{r}$. We assume that each particle consists of a non-magnetic, isotropic and linear material. The assumption of discrete particles enables us to define a distinct scattering source current $\bi{J}_{\rm{S},j}$ in each particle $j$, so that $\bi{J}_{\rm{S}} = \sum_j \bi{J}_{\rm{S},j}$. Then, starting from the ordinary macroscopic Maxwell equations, we derive the following equations to hold for each particle $j$:
\begin{eqnarray}\label{Maxwell1}
\nabla\cdot\bi{E}_{\rm{s},j}(\bi{r}) &= -\frac{\rmi\eta}{k}\nabla\cdot\bi{J}_{\rm{S},j}(\bi{r}),\\\label{Maxwell2}
\nabla\cdot\bi{H}_{\rm{s},j}(\bi{r}) &= 0,\\\label{Maxwell3}
\nabla\times\bi{E}_{\rm{s},j}(\bi{r}) &= \rmi k\eta\bi{H}_{\rm{s},j}(\bi{r}),\\\label{Maxwell4}
\nabla\times\bi{H}_{\rm{s},j}(\bi{r}) &= -\frac{\rmi k}{\eta}\bi{E}_{\rm{s},j}(\bi{r}) + \bi{J}_{\rm{S},j}(\bi{r}).
\end{eqnarray}
In these equations, the incident field and the fields scattered by adjacent particles are implicitly present through $\bi{J}_{\rm{S},j}$. According to (\ref{Maxwell1})-(\ref{Maxwell4}), the introduced $\bi{J}_{\rm{S},j}$ describes the effective current density that creates the scattered field of the $j$'th particle in the self-consistent solution of the Maxwell equations.

Using (\ref{Maxwell1})-(\ref{Maxwell4}), we derive the scalar wave equations \cite{JacksonBook}
\begin{eqnarray}\label{wavE}\fl
\big(\nabla^2 + k^2\big)\big[\bi{r}\cdot\bi{E}_{\rm{s},j}(\bi{r})\big] &= -\rmi k\eta\bi{r}\cdot\bi{J}_{\rm{S},j}(\bi{r}) -\rmi\frac{\eta}{k}\big(2+r\frac{\rmd}{\rmd r}\big)\big[\nabla\cdot\bi{J}_{\rm{S},j}(\bi{r})\big],\\\label{wavH}\fl
\big(\nabla^2 + k^2\big)\big[\bi{r}\cdot\bi{H}_{\rm{s},j}(\bi{r})\big] &= -\bi{r}\cdot\big[\nabla\times\bi{J}_{\rm{S},j}(\bi{r})\big].
\end{eqnarray}
The solutions to (\ref{wavE}) and (\ref{wavH}) are inserted into (\ref{aeSphere}) and (\ref{amSphereAlt}) to obtain the multipole coefficients in the form (see \cite{JacksonBook} for comparison)
\begin{eqnarray}\nonumber\fl
a_{\rm{E}}(l,m) = \frac{(-\rmi)^{l-1}k\eta}{E_0[\pi(2l+1)l(l+1)]^{1/2}}\int Y^*_{lm}(\theta,\phi)j_l(kr)\\\label{aeJS1}
\big\{k^2\bi{r}\cdot\bi{J}_{\rm{S},j}(\bi{r})+\big(2+r\frac{\rmd}{\rmd r}\big)[\nabla\cdot\bi{J}_{\rm{S},j}(\bi{r})]\big\}d^3r,\\\label{amJS1}\fl
a_{\rm{M}}(l,m) = \frac{(-\rmi)^{l-1}k^2\eta}{E_0[\pi(2l+1)l(l+1)]^{1/2}}\int Y^*_{lm}(\theta,\phi)j_l(kr)\bi{r}\cdot[\nabla\times\bi{J}_{\rm{S},j}(\bi{r})]d^3r,
\end{eqnarray}
where $j_l$ are the spherical Bessel functions. While the integrations in (\ref{aeJS1}) and (\ref{amJS1}) are over the whole space, the integrands are equal to zero everywhere outside the particle in question.

The spatial derivatives of $\bi{J}_{\rm{S},j}$ make (\ref{aeJS1}) and (\ref{amJS1}) cumbersome, especially for numerical calculations. We therefore use integration by parts to cast (\ref{aeJS1}) and (\ref{amJS1}) in the form
\begin{eqnarray}\label{aeJ}\fl
a_{\rm{E}}(l,m) = \frac{(-\rmi)^{l-1}k^2\eta O_{lm}}{E_0[\pi(2l+1)]^{1/2}}\int \nonumber \exp{(-\rmi m\phi)}\Big\{\big[\Psi_l(kr)+\Psi_l^{''}(kr)\big]P_l^m(\cos\theta)\hat{\bi{r}}\cdot \bi{J}_{\rm{S},j}(\bi{r}) \\
+ \frac{\Psi_l^{'}(kr)}{kr}\big[\tau_{lm}(\theta)\hat{\bi{\theta}}\cdot\bi{J}_{\rm{S},j}(\bi{r}) - \rmi\pi_{lm}(\theta)\hat{\bi{\phi}}\cdot \bi{J}_{\rm{S},j}(\bi{r})\big]\Big\}d^3r,\\\label{amJ}\fl
a_{\rm{M}}(l,m) = \frac{(-\rmi)^{l+1}k^2\eta O_{lm}}{E_0[\pi(2l+1)]^{1/2}}\int \exp{(-\rmi m\phi)}j_l(kr)\big[\rmi\pi_{lm}(\theta)\hat{\bi{\theta}}\cdot\bi{J}_{\rm{S},j}(\bi{r}) \nonumber\\
+ \tau_{lm}(\theta)\hat{\bi{\phi}}\cdot\bi{J}_{\rm{S},j}(\bi{r})\big]d^3r,
\end{eqnarray}
where $\Psi_l(kr) = krj_l(kr)$ are the Riccati-Bessel functions and $\Psi_l^{'}(kr)$ and $\Psi_l^{''}(kr)$ are their first and second derivatives with respect to the argument $kr$. The associated Legendre polynomials $P_l^m$ are defined as in \cite{JacksonBook}. In (\ref{aeJ}) and (\ref{amJ}) we have introduced the following functions and parameters
\begin{eqnarray}
O_{lm} &= \frac{1}{[l(l+1)]^{1/2}}\Big[\frac{2l+1}{4\pi}\frac{(l-m)!}{(l+m)!}\Big]^{1/2},\\
\tau_{lm}(\theta) &= \frac{d}{d\theta}P_l^{m}(\cos\theta),\\
\pi_{lm}(\theta) &= \frac{m}{\sin\theta}P_l^m(\cos\theta).
\end{eqnarray}

Equations~(\ref{aeJ}) and (\ref{amJ}) yield the same multipole coefficients as (\ref{aeSphere}) and (\ref{amSphere}) for the light scattered by an isolated particle. However, equations~(\ref{aeSphere}) and (\ref{amSphere}) are not applicable to an array of scatterers.
In contrast, since (\ref{aeJ}) and (\ref{amJ}) only require the knowledge of the total electric field inside the particles to calculate $\bi{J}_{\rm{S},j}$, they can be used to characterize the scattering of light by each particle in the array. The required total electric field can be calculated, e.g., by numerically solving the Maxwell equations. Thus, equations~(\ref{aeJ}) and (\ref{amJ}) make it possible to characterize the optical properties of nanomaterials, in which multipoles of arbitrarily high order can be excited.

For scattering of light by a single particle one can introduce the scattering cross section that describes the efficiency with which the particle removes energy from the incident plane wave into the scattered field. For our multipole coefficients the scattering cross section can be derived in a similar way as in \cite{BohrenBook} to become
\begin{equation}\label{Csca}
C_{\rm{s}} = \frac{\pi}{k^2}\sum_{l=1}^{\infty}\sum_{m=-l}^{l} (2l+1)\big[|a_{\rm{E}}(l,m)|^2+|a_{\rm{M}}(l,m)|^2\big].
\end{equation}
The terms of the series in (\ref{Csca}) allow one to determine the contribution of each multipole excitation to the overall scattering cross section of the particle.

The extinction cross section can also be expressed in terms of the multipole coefficients. For an x-polarized incident wave of the form
\begin{equation}
\label{Ex}
\bi{E}_{\rm{i}}(\bi{r}) = \hat{\bi{x}}E_0\exp{(\rmi kz)},
\end{equation}
the extinction cross section is calculated to be given by
\begin{equation}\label{Cext}
C_{\rm{ext,x}} = -\frac{\pi}{k^2}\sum_{l=1}^{\infty}\sum_{m=-1,+1}(2l+1)\mathrm{Re}\big[ma_{\rm{E}}(l,m) + a_{\rm{M}}(l,m)\big].
\end{equation}
In contrast to (\ref{Csca}), the expression for the extinction cross section depends on the choice of the polarization and propagation direction of the incident field. For example, if the incident wave is y-polarized, the extinction cross section is expressed as
\begin{equation}\label{Cexty}
C_{\rm{ext,y}} = \frac{\pi}{k^2}\sum_{l=1}^{\infty}\sum_{m=-1,+1}(2l+1)\mathrm{Im}\big[a_{\rm{E}}(l,m) + ma_{\rm{M}}(l,m)\big].
\end{equation}
The extinction and scattering cross sections can also be calculated by using the optical theorem \cite{JacksonBook} and calculating the total power of the scattered light. The results can then be compared with those obtained by using (\ref{Csca}), (\ref{Cext}) and (\ref{Cexty}), which can serve as an additional check for computations.


\section{The multipole expansion of the electric current density\label{Sec3}}
Scattering of light by a particle can effectively be seen as a process, where the incident light excites in the particle polarization and conduction currents that radiate. These currents can be decomposed into terms which we call current multipoles. To accomplish this decomposition, we choose to use the concept of point electric current elements introduced in \cite{HarringtonBook}. A particle interacting with light can then be treated as a collection of such point elements.

Before presenting the multipole expansion of the electric current density, we first consider a wire of length $L$ that carries a time-harmonic electric current with a complex amplitude $\bi{I}$. The wire is positioned at the origin of the coordinate system inside a homogeneous and isotropic dielectric. The oscillating current emits electromagnetic radiation of wavelength $\lambda$ into the dielectric. By assuming that $L << \lambda$, we can treat this current-carrying wire as a point current element with the complex amplitude of the current density given by
\begin{equation}\label{1element}
\bi{J}_1(\bi{r}) = \bi{I} L\delta(\bi{r}).
\end{equation}
This current element fully corresponds to an oscillating point electric dipole.

\begin{figure}[htb]
\includegraphics{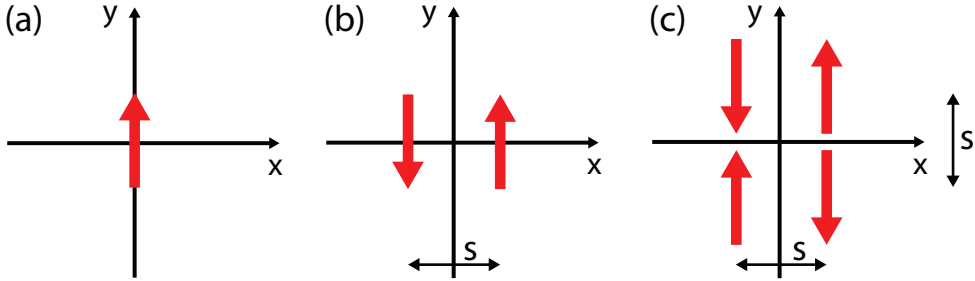}
\caption{Illustration of point current elements (a) $\bi{J}_1 = \hat{\bi{y}}IL\delta(\bi{r})$ , (b) $\bi{J}_2 = \varsigma(\hat{\bi{x}})\bi{J}_1$ and (c) $\bi{J}_3 = \varsigma(\hat{\bi{x}})\varsigma(\hat{\bi{y}})\bi{J}_1$. Each red arrow represents a time-harmonic electric current of complex amplitude $I$ flowing in a wire of length $L$. \label{elementsfig}}
\end{figure}

Next we consider two point elements, in which the currents oscillate in opposite directions. The element with a current $+\bi{I}$ is displaced from the origin in the positive $\hat{\bi{x}}$ direction by a distance $s/2$, while the other element, with a current $-\bi{I}$, is displaced in the opposite direction by the same amount. If $s << \lambda$, we can treat this elementary current configuration as a second-order current element. Considering $s$ to be infinitesimally small, but such that the product $\bi{I} L s$ stays finite, we obtain for the complex amplitude of the current density describing this second-order element
\begin{equation}\label{2element}
\bi{J}_2(\bi{r}) = \varsigma(\hat{\bi{x}})\bi{J}_1(\bi{r}),
\end{equation}
where the operator $\varsigma$ is defined as
\begin{equation}\label{3_multiOper}
\varsigma(\hat{\bi{u}}) = -s\frac{\rmd}{\rmd u}.
\end{equation}
Similar displacements can be done in the $\hat{\bi{y}}$ and $\hat{\bi{z}}$ directions by applying operators $\varsigma(\hat{\bi{y}})$ and $\varsigma(\hat{\bi{z}})$, respectively. The point current elements of the third and higher orders can be obtained by sequentially applying the operator in (\ref{3_multiOper}) to the current density of the lowest order point element of (\ref{1element}). In (\ref{3_multiOper}), $\hat{\bi{u}}$ can be chosen as $\hat{\bi{x}}$, $\hat{\bi{y}}$, $\hat{\bi{z}}$ or as any linear combination of them. The effect of the operator on a point current element of a certain order can be seen as follows. The operator makes a copy of the element and shifts the phase of the complex amplitude of this copy by $\pi$ radians. The original element is then displaced in the $\hat{\bi{u}}$ direction by a distance $s/2$ and the copy in the $-\hat{\bi{u}}$ direction by the same amount (see figure \ref{elementsfig}). Finally, $s$ is set to be infinitesimally small.

We describe the amplitudes of the point current elements by the following $l$'th order current multipole moments
\begin{equation}\label{3_multiMoment}
\dyad{M}^{(l)} = \frac{\rmi}{(l-1)!\omega}\int \bi{J}(\bi{r})\underbrace{\bi{r}\bi{r}...\bi{r}}_{l-1 \mathrm{\:terms}}d^3r,
\end{equation}
where $\dyad{M}^{(l)}$ is a tensor of rank $l$. For the orders $l = 1$, $l = 2$ and $l = 3$, we call the current multipole moments as dipole ($\dyad{M}^{(1)} = \bi{p}$), quadrupole ($\dyad{M}^{(2)} = \dyad{Q}$), and octupole ($\dyad{M}^{(3)} = \dyad{O})$ moments. For $l>2$, several elements of $\dyad{M}^{(l)}$ are equal. For example, the current octupole moments are such that $O_{\rm{xyz}} = O_{\rm{xzy}}$.

The multipole expansion of the current density, obtained by repeatedly applying the operator $\varsigma$ to (\ref{1element}), can be written in Cartesian coordinates as
\begin{eqnarray}\nonumber
\bi{J}(\bi{r}) = &\rmi\omega \sum_{l=1}^{\infty}\sum_{\hat{\bi{v}} = \hat{\bi{x}},\hat{\bi{y}},\hat{\bi{z}}}\sum_{a=0}^{l-1}\sum_{b=0}^{l-a-1}M^{(l)}(\hat{\bi{v}},a,b)\hat{\bi{v}}\\
&\frac{(-1)^{l}(l-1)!}{a!b![l-(a+b+1)]!} \frac{\rmd^a}{{\rmd x}^a}\frac{\rmd^b}{{\rmd y}^b}\frac{\rmd^{l-(a+b+1)}}{{\rmd z}^{l-(a+b+1)}}\delta(\bi{r}),\label{3_expJ}
\end{eqnarray}
where $M^{(l)}(\hat{\bi{v}},a,b)$, with $\hat{\bi{v}}$ being equal to $\hat{\bi{x}}$, $\hat{\bi{y}}$ or $\hat{\bi{z}}$, are the elements of $\dyad{M}^{(l)}$ describing a multipole obtained by applying the operator $\varsigma(\hat{\bi{u}})$ to a $\hat{\bi{v}}$-oriented current element $a$ times with $\hat{\bi{u}} = \hat{\bi{x}}$, $b$ times with $\hat{\bi{u}} = \hat{\bi{y}}$ and $l-(a+b+1)$ times with $\hat{\bi{u}} = \hat{\bi{z}}$. The coefficients in (\ref{3_expJ}) are chosen such that the elements of the multipole tensors are consistent with (\ref{3_multiMoment}).

We wish to map the elements of $\dyad{M}^{(l)}$ onto the coefficients $a_{\rm{E}}(l,m)$ and $a_{\rm{M}}(l,m)$ in the multipole expansion of (\ref{multiE}) and (\ref{multiH}). For this we need to solve for the electromagnetic fields created by the current density distribution of (\ref{3_expJ}). We start by defining the vector potential $\bi{A}$ through
\begin{equation}\label{3_defA}
\bi{H}(\bi{r}) = \frac{1}{\mu_0}\nabla\times \bi{A}(\bi{r}).
\end{equation}
In the Lorenz gauge, $\bi{A}$ satisfies the wave equation
\begin{equation}\label{3_wavA}
\big(\nabla^2+k^2\big)\bi{A}(\bi{r}) = -\mu_0\bi{J}(\bi{r}).
\end{equation}
For a point electric current element, such as the one in (\ref{1element}), the solution for the wave equation is \cite{JacksonBook}
\begin{equation}\label{3_Adipole}
\bi{A}_1(\bi{r}) = \frac{1}{\omega}\frac{k^3\bi{p}}{4\pi\epsilon}h_0^{(1)}(kr),
\end{equation}
where $\bi{p} = \rmi\bi{I}L/\omega$ and $\epsilon = \epsilon_0\epsilon_{\rm{r}}$ is the electric permittivity of the surrounding dielectric. The multipole expansion of the vector potential is obtained by applying the operator $\varsigma$ to (\ref{3_wavA}) with $\bi{J} = \bi{J}_1$ and $\bi{A} = \bi{A}_1$, until $\bi{J}$ on the right-hand-side becomes that in (\ref{3_expJ}). Since the spatial differential operators in Cartesian coordinates commute with $\nabla^2$, we obtain the multipole expansion for $\bi{A}$ from (\ref{3_wavA}) as
\begin{eqnarray}\nonumber
\bi{A}(\bi{r}) = &\frac{1}{\omega}\frac{k^3}{4\pi\epsilon}\sum_{l=1}^{\infty}\sum_{\hat{\bi{v}} = \hat{\bi{x}},\hat{\bi{y}},\hat{\bi{z}}}\sum_{a=0}^{l-1}\sum_{b=0}^{l-a-1}M^{(l)}(\hat{\bi{v}},a,b)\hat{\bi{v}}\\ &\frac{(-1)^{l-1}(l-1)!}{a!b![l-(a+b+1)]!}  \frac{\rmd^a}{{\rmd x}^a}\frac{\rmd^b}{{\rmd y}^b}\frac{\rmd^{l-(a+b+1)}}{{\rmd z}^{l-(a+b+1)}}h_0^{(1)}(kr).\label{Axyz}
\end{eqnarray}

We found that the calculations can be significantly simplified by making use of a circular coordinate system $(w,w^{*},z)$, where $w = (x+\rmi y)/\sqrt{2}$ and $w^{*}$ is its complex conjugate. The unit vectors in this system are $\hat{\bi{w}}$, $\hat{\bi{w}}^{*}$ and $\hat{\bi{z}}$, where $\hat{\bi{w}} = (\hat{\bi{x}}-\rmi \hat{\bi{y}})/\sqrt{2}$. These circular coordinates are linearly related to the Cartesian ones and, therefore, the operator $\varsigma$ in (\ref{3_multiOper}) still commutes with itself. Thus, the vector potential can be expanded as
\begin{eqnarray}\nonumber
\bi{A}(\bi{r}) = &\frac{1}{\omega}\frac{k^3}{4\pi\epsilon}\sum_{l=1}^{\infty}\sum_{\hat{\bi{v}} = \hat{\bi{w}},\hat{\bi{w}}^*,\hat{\bi{z}}}\sum_{a=0}^{l-1}\sum_{b=0}^{l-a-1}M^{'(l)}(\hat{\bi{v}},a,b)\hat{\bi{v}}\\ &\frac{(-1)^{l-1}(l-1)!}{a!b![l-(a+b+1)]!}  \frac{\rmd^a}{{\rmd w}^a}\frac{\rmd^b}{{\rmd w^{*}}^b}\frac{\rmd^{l-(a+b+1)}}{{\rmd z}^{l-(a+b+1)}}h_0^{(1)}(kr),\label{3_expA}
\end{eqnarray}
where $M^{'(l)}(\hat{\bi{v}},a,b)$ are the elements of $\dyad{M}^{(l)}$ in the circular coordinate system.
The magnetic field corresponding to (\ref{3_expA}) is given by (\ref{3_defA}). Outside the scatterer, the electric field is calculated as
\begin{equation}\label{3_EfromA}
\bi{E}(\bi{r}) = \rmi\omega\Big\{\bi{A}(\bi{r}) + \frac{1}{k^2}\nabla\big[\nabla\cdot\bi{A}(\bi{r})\big]\Big\},
\end{equation}
where (\ref{Maxwell4}) and (\ref{3_wavA}) have been used. The multipole coefficients $a_{\rm{E}}(l,m)$ and $a_{\rm{M}}(l,m)$ are obtained by inserting the $\hat{\bi{r}}$ components of $\bi{E}$ and $\bi{H}$ into (\ref{aeSphere}) and (\ref{amSphereAlt}).
For each multipole, the $\hat{\bi{r}}$ components of $\bi{E}$ and $\bi{H}$ can be written in terms of scalar spherical harmonics. The orthogonality of the spherical harmonics can then be used to solve for $a_{\rm{E}}(l,m)$ and $a_{\rm{M}}(l,m)$, without the need to manually calculate the integrals in (\ref{aeSphere}) and (\ref{amSphereAlt}). The elements of the current multipole tensors obtained in the circular coordinate system can be related to the tensor elements in the Cartesian coordinate system by using the tensor projections. In complex coordinates, the projection of the tensor $\dyad{M}^{(l)}$ onto the unit vector $\hat{\bi{w}}$ is $\dyad{M}^{(l)}\cdot\hat{\bi{w}}^{*}$. After lengthy calculations, the following mapping relations are obtained (up to the electric octupole coefficients)
\begin{eqnarray}\label{ae33}
a_{\rm{E}}(3,\pm3) =& \sqrt{15}C_3\big[\pm(O_{\rm{xxx}}-O_{\rm{xyy}}-2O_{\rm{yyx}})\nonumber\\
&+\rmi(O_{\rm{yyy}}-O_{\rm{yxx}}-2O_{\rm{xxy}})\big],\\
a_{\rm{E}}(3,\pm2) =& \sqrt{10}C_3\big[-2O_{\rm{xxz}}+2O_{\rm{yyz}}-O_{\rm{zxx}}+O_{\rm{zyy}}\nonumber\\
&\pm \rmi(2O_{\rm{xyz}}+2O_{\rm{yzx}}+2O_{\rm{zxy}})\big],\\
a_{\rm{E}}(3,\pm1) =& C_3\big[\mp(3O_{\rm{xxx}}+O_{\rm{xyy}}-4O_{\rm{xzz}}+2O_{\rm{yyx}}-8O_{\rm{zzx}})\nonumber\\
&+\rmi(3O_{\rm{yyy}}+O_{\rm{yxx}}-4O_{\rm{yzz}}+2O_{\rm{xxy}}-8O_{\rm{zzy}})],\\
a_{\rm{E}}(3,0) =& 2\sqrt{3}C_3\big[2O_{\rm{xxz}}+2O_{\rm{yyz}}-2O_{\rm{zzz}}+O_{\rm{zxx}}+O_{\rm{zyy}}\big],\\
a_{\rm{E}}(2,\pm2) =& 3C_2\big[Q_{\rm{xx}}-Q_{\rm{yy}}\mp \rmi(Q_{\rm{xy}}+Q_{\rm{yx}})\big],\\
a_{\rm{E}}(2,\pm1) =& 3C_2\big[\mp(Q_{\rm{xz}}+Q_{\rm{zx}})+\rmi(Q_{\rm{yz}}+Q_{\rm{zy}})\big],\\
a_{\rm{E}}(2,0) =& \sqrt{6}C_2\big[2Q_{\rm{zz}}-Q_{\rm{xx}}-Q_{\rm{yy}}\big],\\
a_{\rm{E}}(1,\pm1) =& C_1\big[\mp p_{\rm{x}} +\rmi p_{\rm{y}}\big] \nonumber\\
& +7C_3\big[\pm(O_{\rm{xxx}}+2O_{\rm{xyy}}+2O_{\rm{xzz}}-O_{\rm{yyx}}-O_{\rm{zzx}})\nonumber\\
&-\rmi(O_{\rm{yyy}}+2O_{\rm{yxx}}+2O_{\rm{yzz}}-O_{\rm{xxy}}-O_{\rm{zzy}})\big],\\
a_{\rm{E}}(1,0) =& \sqrt{2}C_1p_{\rm{z}}\nonumber\\
&+7\sqrt{2}C_3\big[O_{\rm{xxz}}+O_{\rm{yyz}}-O_{\rm{zzz}}-2O_{\rm{zxx}}-2O_{\rm{zyy}}\big],\\
a_{\rm{M}}(2,\pm2) =& 7C_3\big[\pm(-O_{\rm{xxz}}+O_{\rm{yyz}}+O_{\rm{zxx}}-O_{\rm{zyy}})\nonumber\\
&+\rmi(O_{\rm{xyz}}+O_{\rm{yzx}}-2O_{\rm{zxy}})\big],\\
a_{\rm{M}}(2,\pm1) =& 7C_3\big[-O_{\rm{xyy}}+O_{\rm{xzz}}+O_{\rm{yyx}}-O_{\rm{zzx}}\nonumber\\
&\mp \rmi(-O_{\rm{yxx}}+O_{\rm{yzz}}+O_{\rm{xxy}}-O_{\rm{zzy}})\big],\\
a_{\rm{M}}(2,0) =& 7\sqrt{6}\rmi C_3\big[O_{\rm{xyz}}-O_{\rm{yzx}}\big],\\
a_{\rm{M}}(1,\pm1) =& 5C_2\big[-Q_{\rm{xz}}+Q_{\rm{zx}}\mp \rmi(-Q_{\rm{yz}}+Q_{\rm{zy}})\big],\\\label{am10}
a_{\rm{M}}(1,0) =& 5\sqrt{2}\rmi C_2\big[-Q_{\rm{xy}}+Q_{\rm{yx}}\big],
\end{eqnarray}
where $C_1 = -\rmi k^3/(6\pi\epsilon E_0)$, $C_2 = -k^4/(60\pi\epsilon E_0)$ and $C_3 = -\rmi k^5/(210\pi\epsilon E_0)$.

We have further verified the correctness of (\ref{ae33})-(\ref{am10}) numerically. We used the computer software COMSOL Multiphysics to numerically calculate the electromagnetic field created by electromagnetic point sources in subwavelength dipole, quadrupole and octupole configurations corresponding to each tensor element in (\ref{ae33})-(\ref{am10}). Using the obtained electric fields in (\ref{aeSphere}) and (\ref{amSphere}), we evaluated the multipole coefficients for each tensor element in $\bi{p}$, $\dyad{Q}$ and $\dyad{O}$. The obtained coefficients were in full agreement with (\ref{ae33})-(\ref{am10}).

The presented theory can be used as follows. For an arbitrary nanoscatterer, including a nanoscatterer in an array, one first numerically evaluates the fields inside the scatterer and uses (\ref{JS}), (\ref{aeJ}) and (\ref{amJ}) to obtain the multipole coefficients $a_{\rm{E}}(l,m)$ and $a_{\rm{M}}(l,m)$. Then (\ref{ae33})-(\ref{am10}) are used to find the essential current excitations in the scatterer. Each tensor element corresponds to a certain electric current mode in the scatterer. For example, the $O_{\rm{xyz}}$ mode describes the current configuration obtained by operating with both $\varsigma(\hat{\bi{y}})$ and $\varsigma(\hat{\bi{z}})$ on an x-oriented current element. Following this recipe, we have, e.g., designed and characterized nanoscatterers in which incident light does not excite any electric dipole moment \cite{Metadimer}.

Let us consider some important properties of the derived current multipole tensors, starting with the quadrupole one. It can be seen that only 8 coefficients [$a_{\rm{E}}(2,\pm2)$, $a_{\rm{E}}(2,\pm1)$, $a_{\rm{E}}(2,0)$, $a_{\rm{M}}(1,\pm1)$ and $a_{\rm{M}}(1,0)$] are used in the multipole expansion, whereas there are 9 elements in the Cartesian current quadrupole dyadic $\dyad{Q}$. This is explained by the fact that the spherically symmetric excitation with $Q_{\rm{xx}} = Q_{\rm{yy}} = Q_{\rm{zz}}$ does not generate any electromagnetic field. This perfectly dark excitation corresponds to a radially oscillating positively charged spherical shell with an equal negative charge at the center. Thus, from the knowledge of the radiated electromagnetic field, the moments $Q_{\rm{xx}}$, $Q_{\rm{yy}}$ and $Q_{\rm{zz}}$ cannot be determined uniquely. However, the real currents should match the geometry of the scatterer, which allows one to make a unique, physically justified choice for the values of $Q_{\rm{xx}}$, $Q_{\rm{yy}}$ and $Q_{\rm{zz}}$. This choice enables one to uniquely specify the excitation character. For the octupoles, there are 15 multipole expansion coefficients (7 electric with $l=3$, 3 electric with $l=1$, and 5 magnetic with $l=2$), but 18 different elements in the Cartesian current octupole tensor. Similarly to spherically symmetric quadrupoles, the three symmetric octupole excitations $2O_{\rm{xxx}}=O_{\rm{yyx}}=O_{\rm{zzx}}$, $O_{\rm{xxy}}=2O_{\rm{yyy}}=O_{\rm{zzy}}$, and $O_{\rm{xxz}}=O_{\rm{yyz}}=2O_{\rm{zzz}}$ are perfectly dark. One of the octupoles in each of these excitations can therefore be chosen arbitrarily.

\begin{figure}[htb]
\includegraphics[width=100mm]{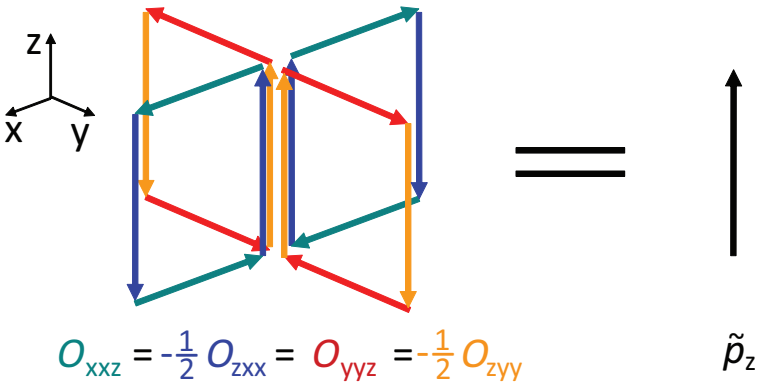}
\caption{The electromagnetic field created by the point octupole on the left is indistinguishable from that created by a point dipole, when $\tilde{p}_{\rm{z}}=2O_{\rm{xxz}}k^2$.\label{octudipole}}
\end{figure}

From (\ref{ae33})-(\ref{am10}) we notice that both current octupoles and dipoles contribute to the same 3 dipole coefficients [$a_{\rm{E}}(1,\pm1)$ and $a_{\rm{E}}(1,0)$] in the multipole expansion. As a consequence, an octupole current distribution \textit{with zero dipole moment} can create an electromagnetic field indistinguishable from an electric dipole. For example, let us consider a current octupole with $\dyad{O} = O(\hat{\bi{x}}\hat{\bi{x}}\hat{\bi{z}}+\hat{\bi{y}}\hat{\bi{y}}\hat{\bi{z}}-2\hat{\bi{z}}\hat{\bi{x}}\hat{\bi{x}}-2\hat{\bi{z}}\hat{\bi{y}}\hat{\bi{y}})$, $\dyad{Q} = 0$ and $\bi{p}=0$. According to (\ref{ae33})-(\ref{am10}), this current distribution creates exactly the same electromagnetic field as a dipole with a moment $\tilde{\bi{p}} = 2Ok^2\hat{\bi{z}}$. This current octupole is illustrated in figure~\ref{octudipole}. The equivalence between the radiation patterns of the current octupole and dipole can also be seen by writing the vector potential in (\ref{Axyz}) for the octupole as
\begin{equation}
\bi{A}(\bi{r}) = \frac{1}{\omega}\frac{k^3}{4\pi\epsilon}2O\big(\nabla\frac{\rmd}{\rmd z}+\hat{\bi{z}}k^2\big)h_0^{(1)}(kr),
\end{equation}
where we have used the fact that $\nabla^2h_0^{(1)}(kr)=-k^2h_0^{(1)}(kr)$. The second term is precisely the vector potential of a dipole with a moment $\tilde{\bi{p}} = 2Ok^2\hat{\bi{z}}$. The first term is a gradient of a scalar function that does not contribute to the radiated field [see (\ref{3_defA})]. This finding emphasizes the importance of the relations in (\ref{ae33})-(\ref{am10}) when using multipole coefficients to describe electromagnetic excitations. The fact that spatially orthogonal current excitations can create the same electromagnetic fields, provides additional freedom to the choice of the particle geometry when designing functional optical nanomaterials.

\section{Conclusions\label{Sec4}}
In summary, we have introduced a theoretical approach to calculate electromagnetic multipole excitations in a material consisting of localized nanostructures in a dielectric host medium. Propagation of light through an array of such nanostructures can be studied numerically by solving the Maxwell equations with appropriate boundary conditions. Then, using our theory, the multipole excitations in the structures can be revealed.

In order to obtain an intuitive picture about the real electric current excitations in the scatterers, we have introduced a basis of easily visualizable electric current multipoles such that any excitation can be represented by their linear superposition. The multipole expansion coefficients can be calculated numerically and used to find the real electric current modes in the nanostructure. The same equations can also be exploited in the reverse order to tailor the angular distribution and directionality of the scattered radiation. The theory presented in this work provides the reader with an exact recipe and with all necessary equations for the design and characterization of nanoscatterers and optical materials composed of them.

\ack
We wish to thank Prof. B J Hoenders, University of Groningen, for helpful discussions. This work was funded by the Academy of Finland (project 134029).

\end{document}